\documentclass[%
 reprint,
 amsmath,amssymb,
 aps,
showkeys
]{revtex4-2}
\usepackage{graphicx}
\usepackage{dcolumn}
\usepackage{bm}
\usepackage{amsmath}

\usepackage[outdir=./]{epstopdf}
\begin{document}

\preprint{APS/123-QED}

\title{Realizations of Measurement Based Quantum Computing}

\author{Swapnil Nitin Shah}
\affiliation{%
Department of Physics, University of Washington, Seattle, WA 98195, USA
}%

\date{\today}

\begin{abstract}
The Measurement Based Quantum Computation (MBQC) model achieves universal quantum computation by employing projective single qubit measurements with classical feedforward on a highly entangled multipartite cluster state \cite{OneWay}. Rapid advances in improving scalability of quantum computing systems have enabled the generation of large cluster states for implementing MBQC on various platforms. This review focuses on three such efforts, each utilizing a different quantum computing technology viz., superconducting qubits \cite{SC-MBQC}, trapped ion qubits \cite{TI-MBQC} and squeezed photon states \cite{Ph-MBQC}. MBQC is being increasingly employed on optical platforms which can generate large entangled resource states but lack the ability to perform deterministic entangling gates \cite{Ph-MBQC,Ph-MBQC_Supp}.    
\end{abstract}

\keywords{Measurement Based Quantum Computation; Cluster States}
\maketitle
  

\section{\label{Intro}Introduction}
The standard circuit model of quantum computation involves initialization of qubits in a known state followed by a sequence of one and two qubit gates as determined by the algorithm. The result is read out by measuring the qubits in a predetermined computational basis. Unlike classical computation, it is not possible to build a general purpose programmable quantum circuit which can perform an arbitrary computation based on an input program state \cite{Nielsen}. The Measurement Based Quantum Computation (MBQC) model employs successive single qubit measurements on a highly entangled multipartite quantum state, i.e. a cluster state in order to perform computation \cite{OneWay}. The unitary operation performed is determined by the eigenbases chosen for these measurements. MBQC allows the possibility of a general purpose programmable quantum computer, given the ability to adapt measurements based on previous outcomes, on a sufficiently large cluster state \cite{MBQC}. As single qubit measurements reduce entanglement of the cluster state, MBQC is also called one-way quantum computing.   

In this paper, we review the MBQC model and its recent implementations using superconducting qubits \cite{SC-MBQC}, trapped ion qubits \cite{TI-MBQC} and spatial modes of squeezed photons \cite{Ph-MBQC}. With rapid advances in utilizing these continuous variable states of squeezed photons to create large cluster states, MBQC is being increasingly employed on these platforms \cite{Ph-MBQC,Ph-MBQC_Supp}.   

\subsection{\label{CS}Cluster States}
\begin{figure}[t] 
	\includegraphics[width=50mm]{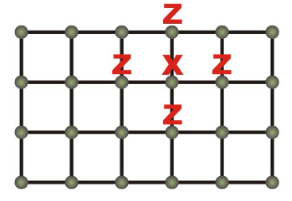}
	\caption{\label{fig1} Visualization of a 2-d cluster state under the action of a stabilizer operator (Source: Ref. \onlinecite{OneWay})}
\end{figure}

Cluster states are multipartite quantum states on a \emph{d}-dimensional lattice of qubits with Ising like nearest neighbor interactions \cite{OneWay}. Representing the neighborhood of a qubit $c$ in a lattice $C$ by $\Gamma_c$, the cluster state $\left|\psi_C\right\rangle$ is given by \cite{OneWay}
\begin{equation}
\left|\psi_C\right\rangle \equiv \mathcal{N}\bigotimes_{c\in C} \left(\left|1\right\rangle_c\bigotimes_{x\in \Gamma_c} Z^{(x)}+\left|0\right\rangle_c\right)
\label{eq_cs}
\end{equation}
where $\mathcal{N}$ is the normalization factor and $Z^{(x)}$ is the Pauli-Z operator applied to qubit $x$. $\left|\psi_C\right\rangle$ is an eigenstate of the stabilizer operator
\begin{equation}
K_a \equiv X_a \bigotimes_{x\in \Gamma_a} Z^{(x)}
\label{eq_sb}
\end{equation} 
where $X_a$ is the Pauli-X operator applied to qubit $a$. Such a state can be realized by performing controlled-Z gates between qubits initialized in the $\left|+\right\rangle \equiv \frac{1}{\sqrt{2}}\left(\left|0\right\rangle+\left|1\right\rangle\right)$ states. Fig. \ref{fig1} shows a 2-d cluster state on which a stabilizer operator is applied.

\subsection{\label{Csch}MBQC Computation Scheme}  
\begin{figure}[hb] 
	\includegraphics[width=68mm]{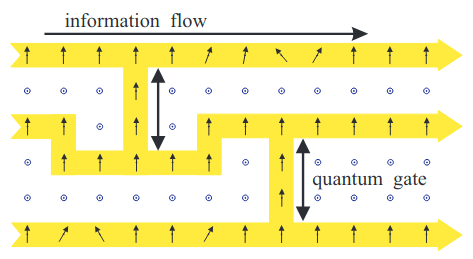}
	\caption{\label{fig2} One-way quantum computation on a cluster state. $\odot$ represents unused qubits measured in the computational (Pauli-Z) basis, $\uparrow$ are qubits measured in Pauli-X basis and tilted arrows are qubits measured in X-Y plane (Source: Ref. \onlinecite{OneWay})}
\end{figure}

Any unitary operation on an array of qubits can be decomposed into one-qubit rotations and universal two-qubit gates (e.g. Controlled-NOT (CNOT)). Once a sufficiently large cluster state (Eq. (\ref{eq_cs})) is available, it can be used for arbitrary unitary operations using such decomposition \cite{OneWay,MBQC}. Measurements implementing the gates are performed in a sequence (imposed by dependence on previous measurements) along one dimension of the cluster state. Individual gates can be carved out in the cluster by measuring the unused qubits in the computational basis \cite{OneWay} (Fig. \ref{fig2}). Such a measurement destroys entanglement between the measured qubit and its neighbors, effectively removing it from the cluster. Formally,
\begin{equation}
P^{(c^*)}_Z\left|\psi_C\right\rangle = \mathcal{N}'\bigotimes_{c\,\in\, C\backslash\{c^*\}} \left(\left|1\right\rangle_c\bigotimes_{x\in \Gamma_c} Z^{(x)}+\left|0\right\rangle_c\right)
\label{eq_carve}
\end{equation}
where $P^{(c^*)}_Z$ is the projection operator in the computational basis for qubit $c^*$. 
The qubit states are transported through the cluster via teleportation which is performed by measurement of qubit state in the Pauli-X ($\left|+\right\rangle,\left|-\right\rangle$) basis. To illustrate, consider Pauli-X measurement of a qubit in state $\left|\psi_{\mathrm{in}}\right\rangle \equiv \alpha\left|0\right\rangle+\beta\left|1\right\rangle$ ($\left|\psi'_{\mathrm{in}}\right\rangle \equiv Z \left|\psi_{\mathrm{in}}\right\rangle= \alpha\left|0\right\rangle-\beta\left|1\right\rangle$) entangled with another qubit in a cluster state
\begin{subequations} 
\begin{equation}
\left|\phi\right\rangle\equiv\left(\frac{1}{\sqrt{2}}\left|\psi_{\mathrm{in}}\right\rangle\otimes\left|0\right\rangle+\frac{1}{\sqrt{2}}\left|\psi'_{\mathrm{in}}\right\rangle\otimes\left|1\right\rangle\right)
\end{equation}
\begin{equation}
P^{(\psi)}_X \left|\phi\right\rangle = \frac{1}{\sqrt{2}} \left(\alpha\left|+\right\rangle+(-1)^{s_1}\beta\left|-\right\rangle\right)
\end{equation}
\end{subequations}
where $s_1=\{0,1\}$ is the outcome of measurement. The state $\left|\psi_{\mathrm{in}}\right\rangle$ (up to a phase correction) is teleported to the other qubit in the Pauli-X basis as a result of measurement \cite{OneWay}. In general, the source qubit state is teleported to the target in the Pauli-X basis for odd number of teleportation steps and in the Pauli-Z basis for even number of steps.

\subsubsection{\label{Gates}Unitary Gates in MBQC}
\begin{figure}[h] 
	\includegraphics[width=50mm]{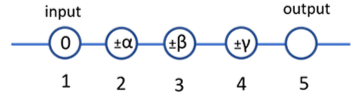}
	\caption{\label{fig3} One qubit rotation using a linear chain of 5 qubits in a cluster state. (Source: Ref. \onlinecite{MBQC})}
\end{figure}

Every one-qubit rotation $U_R\in$ SU(2) admits a decomposition in Euler angles as \cite{MBQC}
\begin{equation}
U_R(\alpha,\beta,\gamma)=U_X(\gamma)U_Z(\beta)U_X(\alpha)
\end{equation}
where $U_X$ and $U_Z$ describe rotations about the Pauli-X and Pauli-Z axis respectively. This can be implemented using a linear chain of 5 qubits on a cluster state with input state on qubit 1 (Fig. \ref{fig3}). Qubits 1$\rightarrow$4 are measured in the following bases \cite{OneWay}
\begin{equation}
\mathcal{B}_j\left(\theta_j\right)\equiv\left\{\frac{\left|0\right\rangle+e^{i\theta_j}\left|1\right\rangle}{\sqrt{2}},\frac{\left|0\right\rangle-e^{i\theta_j}\left|1\right\rangle}{\sqrt{2}}\right\}
\end{equation} 
\begin{equation}
\nonumber
\begin{split}
&\theta_1=0;\;\;\;\theta_2=\alpha(-1)^{s1+1};\\
&\theta_3=\beta(-1)^{s2};\;\;\;\theta_4=\gamma(-1)^{s1+s3}
\end{split}
\end{equation}
where $s1\dots s4$ are the measurement outcomes. The implemented unitary operation is given by $U_g=U_\Sigma U_R$ where $U_\Sigma\equiv X^{s2+s4}Z^{s1+s3}$ describes the correction required based on measurement outcomes.  

\begin{figure}[h] 
	\includegraphics[width=45mm]{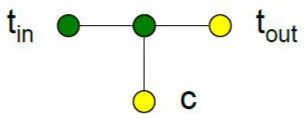}
	\caption{\label{fig4} Minimal CNOT gate implementation on a cluster state. $t_{in}$ is the input, $t_{out}$ is the output and $c$ is the control qubit (Source: Ref. \onlinecite{OneWay})}
\end{figure}

In order to have a universal set of gates, a universal 2-qubit gate is necessary. Fig. \ref{fig4} shows a minimal CNOT implementation on a cluster state. The input qubit $t_{in}$ and its neighbor are measured in the Pauli-X basis, thereby implementing a CNOT (up to Pauli corrections) between control qubit $c$ and output qubit $t_{out}$ \cite{OneWay}. In practice, the control and target qubits are teleported to other qubits (Eq. (4)) in the cluster after computation.
A composite circuit such as in Fig. 2 can be created on the cluster which uses arbitrary one qubit rotations and CNOT gates to perform complex unitary operations on the input qubits.

\section{\label{Imp}Implementations}
Physical realizations of MBQC require the ability to a) Create large cluster states with a high degree of persistent entanglement between the qubits and b) Perform projective measurements on individual qubits forming the cluster in chosen eigenbases with feedforward control \cite{MBQC}. There have been many recent implementations of MBQC model, of which we focus on those employing superconducting qubits \cite{SC-MBQC}, trapped ion qubits \cite{TI-MBQC} and spatio-temporal modes of squeezed photons \cite{Ph-MBQC}. This computation model is especially suitable for photonic quantum systems which are quite scalable but lack the ability to directly perform deterministic entangling gates. In the next few sections, we review recent efforts of implementing MBQC on the aforementioned platforms.

\subsection{\label{Imp-Ph}MBQC with Squeezed Photon Modes}
One can encode quantum information in spatial and temporal modes of squeezed photon states viz., qumodes \cite{GKP}, which are continuous variable analogues of qubits. The Pauli-Z basis is replaced by the position quadrature basis $\left|q\right\rangle_{q\in \mathbb{R}}$ and the Pauli-X basis is replaced by the momentum quadrature basis $\left|p\right\rangle_{p\in \mathbb{R}}$ in this analogue. In the article [Ref. \onlinecite{Ph-MBQC}], the authors utilize squeezed photon states in two spatial modes and standard linear optical devices to implement a) A 2-d Gaussian photonic cluster state with a width of 6 qubits and infinite depth and b) Universal set of one and two qubit Gaussian unitaries using this cluster state.

\subsubsection{\label{PhCS}Continuous Variable Cluster State}
\begin{figure}[h] 
	\includegraphics[width=90mm]{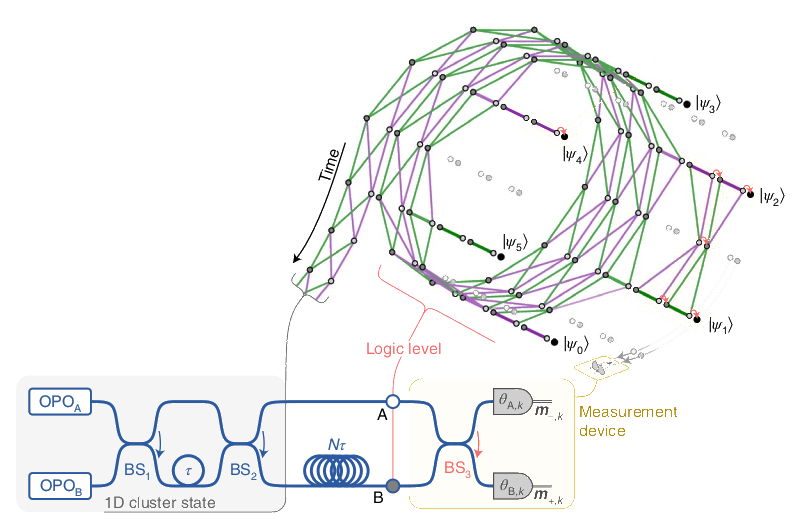}
	\caption{\label{fig5} Experimental setup and visualization of the cylindrical 2-d cluster state. $N=12$ which provides 6 parallel modes for computation (other 6 are control modes). The measurement device consists of a beam splitter and two homodyne detectors measuring in basis $\theta_{A,k},\theta_{B,k}$ for temporal mode $k$ (Source: Ref. \onlinecite{Ph-MBQC})}
\end{figure}

The article [Ref. \onlinecite{Ph-MBQC}] describes generation of a 2-d cluster state with temporal mode encoding of squeezed photon states in two spatial modes, $A$ and $B$ in Fig. \ref{fig5}. Optical Parametric Oscillators (OPO) are used to generate the squeezed photon states which are interfered in an unbalanced interferometer (delay of $\tau$ in spatial mode B) to generate a 1-d cluster state, also known as a dual rail quantum wire \cite{Ph-MBQC_Supp}. Each point in this cluster corresponds to a single spatial mode $s\in\{A,B\}$ and temporal mode $k\in\mathbb{N}$. Using a delay line of $N\tau$ for mode $B$ of the 1-d cluster, it is coiled up into a 2-d cylindrical cluster state. High efficiency homodyne detectors are employed for both spatial modes which can make projective measurements in arbitrary quadrature basis $\hat{q}\cos{\theta}+\hat{p}\sin{\theta}$.  
In Fig. \ref{fig5}, the gray nodes have $k$ odd and are measured in the quadrature basis $\theta_c=(-1)^{(k-1)/2}\pi/4$. This projects the cluster as wires for state teleportation along the length of the 2-d cylinder \cite{Ph-MBQC,Ph-MBQC_Supp}. These form the logic levels for computation of arbitrary multimode gaussian unitaries. If Gottesman-Kitaev-Preskill (GKP) encoded qubits are available, universal fault-tolerant quantum computation is possible with this framework \cite{GKP,Ph-MBQC}.

\subsubsection{\label{PhSMU}Single Mode Gaussian Gates}
\begin{figure}[h] 
	\includegraphics[width=82mm]{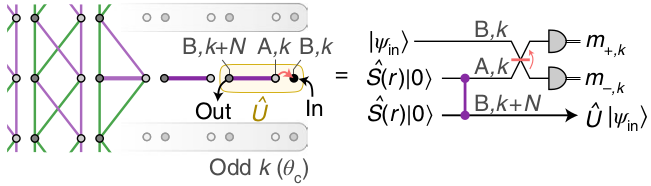}
	\caption{\label{fig6} (Left) Single mode Gaussian gate on a projected wire, (Right) Equivalent teleportation based circuit (Source: Ref. \onlinecite{Ph-MBQC})}
\end{figure}

A single mode Gaussian unitary is implemented using teleportation on a logic level of the 2-d cluster \cite{Ph-MBQC} (Fig. \ref{fig6}). 
The unitary that is implemented depends on the measurement bases for the two homodyne detectors (up to a measurement dependent displacement) as \cite{Ph-MBQC}
\begin{equation}
\hat{U}=(-1)^w\hat{R}\left(\theta_+\right)\hat{S}\left(\tan{\theta_-}\right)\hat{R}\left(\theta_+\right)
\end{equation}  
where $w$ is the wire number, $\theta_\pm=\theta_{A,k}\pm\theta_{B,k}$ and $\hat{R}\left(\theta\right)$, $\hat{S}(t)$ are the quadrature rotation and squeezing unitaries given by
\begin{equation}
\hat{R}\left(\theta\right)\equiv e^{i\theta(\hat{q}^2+\hat{p}^2)/2}, \;\;\;\; \hat{S}(t)\equiv e^{i\ln{t}(\hat{x}\hat{p}+\hat{p}\hat{x})/2}
\end{equation} 
A cascade of two such unitaries (Eq. (7)) can implement any single mode Gaussian unitary (up to a displacement) by choosing appropriate quadrature measurement bases \cite{Ph-MBQC_Supp}. 

\subsubsection{\label{PhTMZ}Two Mode Controlled Phase Gate}
\begin{figure}[h] 
	\includegraphics[width=82mm]{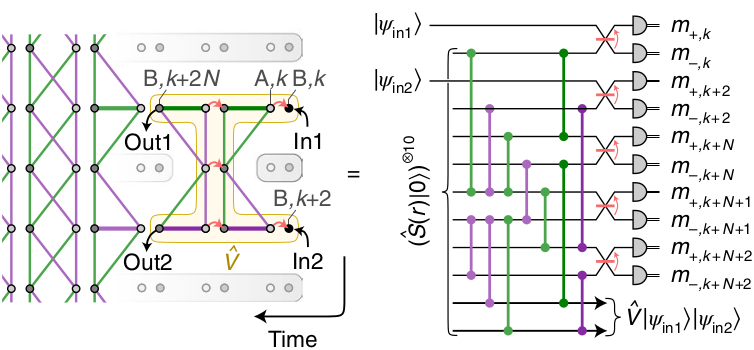}
	\caption{\label{fig7} (Left) Two mode Gaussian gate using two logic levels and a pair of control modes (Right) Equivalent teleportation based circuit (Source: Ref. \onlinecite{Ph-MBQC})}
\end{figure}

To complete the universal set of Gaussian unitaries, an entangling two mode gate must be implemented. The authors demonstrate how entanglement between two logic levels and one pair of control modes can be used to implement a Gaussian unitary with two input modes \cite{Ph-MBQC} (Fig. \ref{fig7}).    
As with the single mode case, the implemented two mode unitary is determined by the measurement bases of the homodyne detectors for each temporal mode. The two mode controlled phase gate $\hat{C}_z\equiv e^{ig(\hat{q}_1\otimes\hat{q}_2)}$ along with the single mode rotation and sqeezing gates (Eq. (8)) form a universal multimode Gaussian gate set \cite{Ph-MBQC}. To implement the controlled phase gate (up to known rotations) between the two input modes, choice of measurement bases are determined to be \cite{Ph-MBQC,Ph-MBQC_Supp}
\begin{equation}
\begin{pmatrix}
\theta_{A,k}\\
\theta_{B,k}\\
\theta_{A,k+2}\\
\theta_{B,k+2}\\
\theta_{A,k+N}\\
\theta_{B,k+N}\\
\theta_{A,k+N+1}\\
\theta_{B,k+N+1}\\
\theta_{A,k+N+2}\\
\theta_{B,k+N+2}
\end{pmatrix}=\begin{pmatrix}
\pi/4\\
-\pi/4\\
(-1)^w\pi/4\\
-(-1)^w\pi/4\\
(-1)^w[\pi/2-\arctan{(g/2)}]\\
0\\
(-1)^w\pi/4\\
(-1)^w[\pi/4+2\arctan{(g/2)}]\\
(-1)^w[\pi/2-\arctan{(g/2)}]\\
0
\end{pmatrix}
\end{equation}
where $w$ is the wire number of the input mode $B,k$ and the implemented unitary transformation is $\left[\hat{R}(\pi/2)\otimes\hat{R}\left((-1)^w\pi/2\right)\right]\hat{C}_z$.   

The above gate implementations assume infinite squeezing, which would require infinite energy and are therefore unphysical. In practice, one employs finitely squeezed photon states which introduce gate noise in the computation which accumulates as more gates are added. For one and two mode gates, the measured gate noise (after accounting for the gate noise factor of 6 dB) agrees well with the measured squeezing variance of -4.4 dB in the initial momentum quadratures. In order to make this approach scalable, some form of error correction would be required. GKP encoded qubits \cite{GKP} (albeit not yet experimentally realized in optical spectrum) is one approach which could be used for fault-tolerant quantum computation with this scheme.

\subsection{\label{Imp-SC}MBQC with Superconducting Qubits}
In the article [Ref. \onlinecite{SC-MBQC}], the authors propose an experimentally feasible implementation of a cluster state of the form - Eq. (1) using high fidelity $\hat{C}_z$ gates on a 2-d lattice of superconducting qubits. They discuss the construction of a universal CNOT gate (Fig. \ref{fig4}) on this plaform and its expected fidelity. An alternative entangling $\hat{U}_{Bell}$ gate is proposed which reduces the number of qubits required for CNOT gate implementation. The 2-d cluster state generated with $\hat{U}_{Bell}$ gates is expected to be Maximally Persistent and Maximally Connected (MPMC) which could potentially lead to increased scalability of such implementations \cite{SC-MBQC}.

\subsubsection{\label{CS-Xmon}Cluster State with Transmon Qubits}
\begin{figure}[h] 
	\includegraphics[width=50mm]{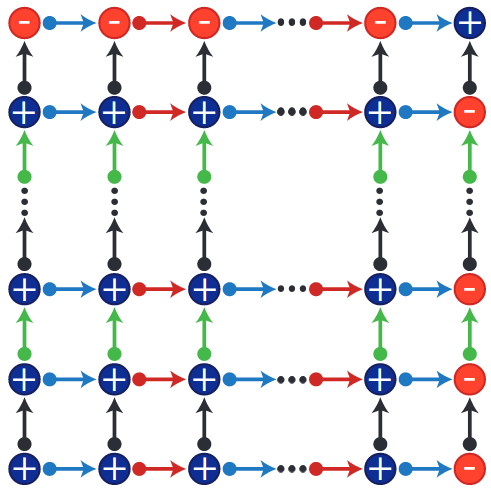}
	\caption{\label{fig8} Cluster state with Transmon qubits. Arrows represent $\hat{C}_z$ gates from control to target. All arrows of same color represent gates that can be applied simultaneously. ``+" and ``-" represent respective state initializations (Source: Ref. \onlinecite{SC-MBQC})}
\end{figure}

The interaction Hamiltonian of a pair of Transmon qubits with same energy gap between ground $\left|0\right\rangle$ and excited $\left|1\right\rangle$ states is of the form \cite{SC-MBQC}
\begin{equation}
\hat{H}_\mathrm{int}=\hbar\, g_\mathrm{eff}\sum_{(j,k),\, j<k} \left(\frac{\mathbb{I}-\sigma^j_z}{2}\right)\left(\frac{\mathbb{I}+\sigma^k_z}{2}\right)
\end{equation} 
For a 2-d lattice of Transmons, time evolution of interaction is given by $\hat{\mathcal{U}}=e^{-i\hat{H}_\mathrm{int}t/\hbar}$. For $g_\mathrm{eff}t=\pi$ and initial states $\left|+\right\rangle$, this describes a cluster state of the form
\begin{equation}
\left|\psi'_C\right\rangle \equiv \mathcal{N}\bigotimes_{c\in C} \left(\left|1\right\rangle_c\bigotimes_{x\in \Gamma_c} -Z^{(x)}+\left|0\right\rangle_c\right)
\end{equation}
which differs in sign of Pauli-Z operator from Eq. (1). To express this cluster state in terms of $\hat{C}_z$ gates, one considers initial state of a qubit to be $\left|+\right\rangle$ for control of even number of qubits and $\left|-\right\rangle$ for control of odd number of qubits (Fig. \ref{fig8}). For a 2-d lattice of Transmon qubits of size $h\times l$, each qubit can be represented by a vector $\vec{p}\equiv(j,k)$, where $(j,k)$ is its coordinate on the lattice \cite{SC-MBQC}. 
Using this, the cluster (Eq. (11)) is given by 
\begin{equation}
\begin{aligned}
\left|\psi'_C\right\rangle=\left[\prod_{\vec{p}\in \left\{\mathcal{L}+\mathcal{L}_x\right\}}\mathcal{C}^{\left(\vec{p},\vec{p}+\hat{m}\right)}_z\right]\left[\prod_{\vec{p}\in \left\{\mathcal{L}+\mathcal{L}_y\right\}}\mathcal{C}^{\left(\vec{p},\vec{p}+\hat{n}\right)}_z\right]\\
\times\left[\bigotimes_{\vec{p}\in \mathcal{L}}\left|+_{\vec{p}}\right\rangle\right]
\left[\bigotimes_{\vec{p}\in \mathcal{L}_x}\left|-_{\vec{p}}\right\rangle\right]\left[\bigotimes_{\vec{p}\in \mathcal{L}_y}\left|-_{\vec{p}}\right\rangle\right]\\
\otimes\left|+_{(h,l)}\right\rangle
\end{aligned}
\end{equation}  
where $\mathcal{L}\equiv\{(j,k)\}$, $\mathcal{L}_x\equiv\{(j,l)\}$, $\mathcal{L}_y\equiv\{(h,k)\}$, $j=\{1,2\dots h-1\}$, $k=\{1,2\dots l-1\}$ and $\hat{m}\equiv(1,0)$, $\hat{n}\equiv(0,1)$.

\subsubsection{\label{CNOT-Xmon}CNOT with Transmon Cluster States}
Implementation of a CNOT gate using the Transmon cluster state (Eq. (11)) follows the recipe shown in Fig. \ref{fig4}. The initial state of the four qubits is
\begin{equation}
\left|\Psi_0\right\rangle\equiv \left|i_1\right\rangle\left|+\right\rangle\left|+\right\rangle\left|c_4\right\rangle
\end{equation}
On applying the three $\hat{C}_z$ gates followed by measurement of qubits 1 and 2 in the Pauli-X basis, this state evolves into \cite{SC-MBQC}
\begin{equation}
\left|\Psi_m\right\rangle\equiv \mathcal{F}_{3,4}\left|l\right\rangle\left|m\right\rangle\left|i\oplus c\right\rangle\left|c_4\right\rangle
\end{equation}
where $l,m$ are measurement outcomes of the respective qubits and $\mathcal{F}_{3,4}\equiv \left(\sigma^3_x\right)^m\left(\sigma^3_z\sigma^4_z\right)^l$ is the measurement based correction. The authors report a $\hat{C}_z$ gate fidelity of $99.5\%$ with an evolution time of $\approx 0.05\mu s$ per gate. Also, the readout and measurement process is estimated to have an error of $2\%$ and a processing time of $\approx 2\mu s$ per measurement. Estimated CNOT gate fidelity is therefore $(0.995^3 - 0.04)\times 100\approx 94.5\%$ with an evaluation time of $<5\mu s$, which is much smaller than Transmon coherence times \cite{SC-MBQC}. 

\begin{figure}[h] 
	\includegraphics[width=45mm]{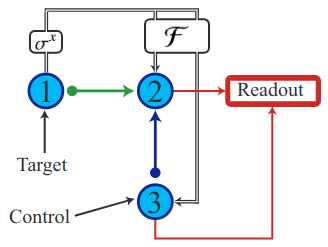}
	\caption{\label{fig9} CNOT using a 3 qubit cluster state. Qubits $1\leftrightarrow 2$ are entangled with a $\hat{U}_{Bell}$ gate and qubits $2\leftrightarrow 3$ are entangled with a $\hat{C}_z$ gate (Source: Ref. \onlinecite{SC-MBQC})}
\end{figure}

The authors further discuss an alternative entangling gate $\hat{U}_{Bell}$ for generating cluster states. The action of this gate is given in block matrix form as
\begin{equation}
\hat{U}_{Bell}\equiv \frac{1}{\sqrt{2}}\begin{bmatrix}
\mathbb{I} & \sigma_x\\
-\sigma_x & \mathbb{I}\\
\end{bmatrix}
\end{equation}
With a cluster state made using a $\hat{U}_{Bell}$ and a $\hat{C}_z$ gate, CNOT can be implemented with only 3 qubits (Fig. \ref{fig9}).
Qubits 1 and 2 are entangled with the $\hat{U}_{Bell}$ gate (Eq. (15)) and qubits 2 and 3 are entangled with a $\hat{C}_z$ gate to form a 3 qubit cluster state to implement CNOT. The initial state of the qubits is
\begin{equation}
\left|\Psi_0\right\rangle\equiv \left|i_1\right\rangle\left|0\right\rangle\left|c_3\right\rangle
\end{equation}
On applying the $\hat{U}_{Bell}$ and $\hat{C}_z$ gates, this evolves into
\begin{equation}
\left|\Psi_e\right\rangle=\left[\left|i_1\right\rangle\left|0\right\rangle+(-1)^{i+c}\left|i\oplus 1\right\rangle\left|1\right\rangle\right]\otimes\left|c_3\right\rangle
\end{equation}
Measurement of qubit 1 in Pauli-X basis followed by Hadamard operation on qubit 2 yields
\begin{equation}
\left|\Psi_m\right\rangle=\mathcal{F}_{2,3}\left|m\right\rangle\left|i\oplus c\right\rangle\left|c_3\right\rangle
\end{equation}
where $m$ is the measurement outcome and $\mathcal{F}_{2,3}\equiv \left(\sigma^3_z\sigma^2_x\sigma^2_z\right)^m$ is the measurement based correction. The authors estimate the fidelity of this CNOT to be $\approx 96.5\%$ for an evaluation time of $2.5\mu s$ which is a significant improvement over the 4 qubit CNOT (Eq. (14)) with a $25\%$ reduction in number of qubits. Beyond CNOT, the cluster state created using only $\hat{U}_{Bell}$ gates has other useful characteristics of being - a) Maximally connected (measuring all but 2 qubits of the cluster results in a Bell state) and b) Maximally persistent (minimal product state decomposition has $2^{N}$ terms for an N qubit cluster state) \cite{SC-MBQC}. 

\subsection{\label{Imp-TI}MBQC with Trapped Ions}
In the article [Ref. \onlinecite{TI-MBQC}], authors use strings of $^{40}\mathrm{Ca}^+$ ions in a linear Paul trap to create one and two dimensional cluster states. They employ MBQC on these cluster states to demonstrate universal one and two qubit gates. Beyond quantum computation, they also demonstrate Quantum Error Correction (QEC) on the generated cluster states using MBQC. 

\subsubsection{\label{CS-TI}Cluster States with Trapped Ions}
The electron states $\left|D_{5/2},m=+3/2\right\rangle$ and $\left|S_{1/2},m=+1/2\right\rangle$ of $^{40}\mathrm{Ca}^+$ ions are used as qubit states $\left|1\right\rangle$ and $\left|0\right\rangle$ respectively ($\lambda=729$ nm) \cite{TI-MBQC}. The $n$ ionic qubits in a linear Paul trap are initialized in state $\left|1\right\rangle^{\otimes n}$ via optical pumping. The axial center of mass and stretch vibrational modes are initialized in ground state using resolved sideband cooling. The cluster states are created from the ion string using three types of interactions - a) Molmer Sorenson interaction which is a long range qubit-qubit interaction of the form $M(\theta)\equiv\exp{\left(-i\theta\sum_{a<b}X_aX_b\right)}$, b) Single qubit phase flips $G_k(\theta)\equiv\exp{\left(-i\frac{\theta}{2}Z_k\right)}$ and c) Hiding pulses (in this case, hiding qubit state $\left|1\right\rangle$ in a different Zeeman level $\left|D_{5/2},m=+5/2\right\rangle$) of the form $F_h\equiv\exp{\left(-i\pi/4\sum_kX_k\right)}$. 
\begin{figure}[h] 
	\includegraphics[width=80mm]{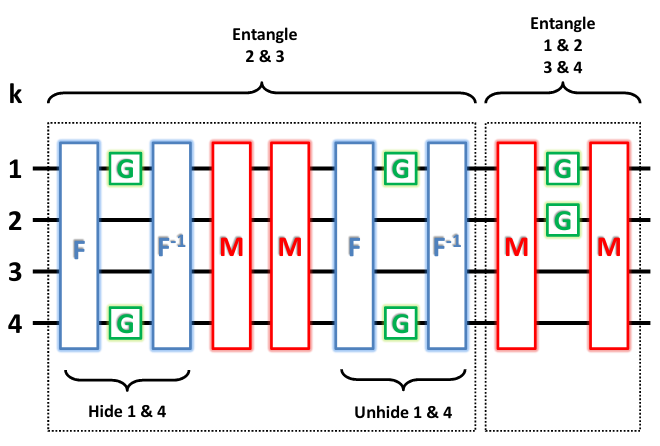}
	\caption{\label{fig10} Quantum circuit to generate a 1-d cluster state with 4 trapped ion qubits using the three types of interactions $(M,G,F)$. (Source: Ref. \onlinecite{TI-MBQC})}
\end{figure}  
Fig. \ref{fig10} \cite{TI-MBQC} illustrates an equivalent quantum circuit to generate a 4 qubit 1-d cluster state using these interactions as gates. The resultant cluster state, $\left|E_{LC_4}\right\rangle$, is equivalent to $\left|{LC_4}\right\rangle$ (described by Eq. (1)) up to single qubit unitary operations.  
\begin{equation}
\begin{split}
\left|E_{LC_4}\right\rangle\equiv \frac{1}{\sqrt{8}} (&\left|0000\right\rangle-i\left|0011\right\rangle-\left|0101\right\rangle\\
-i&\left|0110\right\rangle+i\left|1001\right\rangle-\left|1010\right\rangle\\
-i&\left|1100\right\rangle-\left|1111\right\rangle)
\end{split}
\end{equation}
On performing quantum tomography of the cluster state density matrix, authors found its fidelity with respect to ideal state as being $0.841\pm 0.006$.

\subsubsection{\label{TI-Gates}Universal Gates with Trapped Ion Cluster States}
For implementation of single qubit gates, decomposition into Euler angles and measurements, described by Eq. (5) and (6), is employed with the 1-d cluster state $\left|E_{LC_n}\right\rangle$. Using quantum state tomography, the authors obtained an average fidelity of $0.92\pm 0.01$ for single qubit gates.

\begin{figure}[h] 
	\includegraphics[width=40mm]{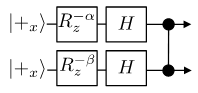}
	\caption{\label{fig11} Two qubit quantum gate implemented using a $\left|{LC_4}\right\rangle$ cluster state. (Source: Ref. \onlinecite{TI-MBQC})}
\end{figure}

For illustration of two qubit gates, qubits 1 and 4 of $\left|{LC_4}\right\rangle$ act as inputs which are initially in state $\left|++\right\rangle$. Measuring them in bases $\mathcal{B}(\alpha)$ and $\mathcal{B}(\beta)$ (Eq. (6)) results in their rotations about the Pauli-Z axis, followed by Hadamard transformation. These are subjected to the teleported $\hat{C}_z$ gate and results are read out from qubits 2 and 3 (Fig. \ref{fig11}). The measurement based correction for this gate is $\mathcal{F}=(\sigma^2_x\sigma^3_z)^{s1}(\sigma^3_x\sigma^2_z)^{s4}$, where $s1$ and $s4$ are the measurement outcomes. Using this scheme, an entangled state $(\alpha=\pi/2, \beta=-\pi/2)$ and a separable state $(\alpha=0, \beta=0)$ were generated. Quantum state tomography yielded respective fidelities of $0.88\pm 0.02$ and $0.83\pm 0.01$ \cite{TI-MBQC}. 
These implementations of a $\hat{C}_z$ gate and single qubit rotations form a universal gate set for quantum computation.

\subsubsection{\label{TI-EC}Quantum Error Correction with Cluster States}
\begin{figure}[h] 
	\includegraphics[width=25mm]{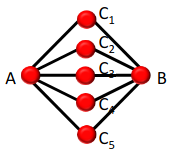}
	\caption{\label{fig12} $\left|EC_5\right\rangle$ cluster state for implementing $5$-qubit phase flip code. The state of qubit $A$ gets encoded in a code-word formed by qubits $C_1\dots C_5$, which can be error corrected at the output qubit $B$ based on the error syndrome. (Source: Ref. \onlinecite{TI-MBQC})}
\end{figure}

Using cluster states, it is possible to implement a phase flip code capable of correcting full phase flips of $(n-1)/2$ of qubits forming the $n$-qubit code-word ($n$ is odd) \cite{TI-MBQC}. The cluster state employed for error correction (Fig. \ref{fig12}) has the form
\begin{equation}
\begin{split}
\left|EC_n\right\rangle\equiv & \left(\left|0\right\rangle_A\left|+\right\rangle^{\otimes n}+\left|1\right\rangle_A\left|-\right\rangle^{\otimes n}\right)\left|0\right\rangle_B\\
+ & \left(\left|0\right\rangle_A\left|-\right\rangle^{\otimes n}+\left|1\right\rangle_A\left|+\right\rangle^{\otimes n}\right)\left|1\right\rangle_B 
\end{split}
\end{equation}

The input qubit $A$ is measured in the eigenbasis of the state to be encoded. This is followed by a measurement of code-word qubits in the Pauli-X basis which both a) teleports the encoded state to the output qubit $B$ and b) reveals the error syndrome for correction of any phase flip errors before measurement. From Eq. (20), it can be seen that if $>(n-1)/2$ of code-word qubits are measured in $\left|+\right\rangle$ state, no recovery operation is required. If, however, $>(n-1)/2$ of code-word qubits are measured in $\left|-\right\rangle$ state, the output state needs to be corrected by applying a $\sigma^B_x$ operation on qubit $B$.

\section{\label{Disc}Discussion}
MBQC achieves universal quantum computation through projective single qubit measurements on a highly entangled cluster state. This simplicity of computation comes at the expense of a large number of qubits forming the cluster state and their coherent control. Trapped ion qubits have the highest reported fidelities ($\approx 99\%$) for single and two qubit gates \cite{TI-Fid}. However, controlling a large number of such qubits in a single trap while maintaining coherence is extremely challenging. The most number of trapped ion qubits that have been coherently controlled so far are $\approx 20$ \cite{TI-20} which can implement a maximum of 5 CNOT gates on a cluster state. Superconducting qubits have reasonably good gate fidelities ($\approx 96\%$) \cite{SC-MBQC} and have better scalability ($\approx 127$ qubits \cite{IBM-127}) than trapped ion qubits. While this may be sufficient for proof of concept algorithms, it is far from being useful in creating large enough cluster states for real world computations. Also, larger superconducting quantum systems require some form of error correction to reduce propagation of noise and effects of decoherence. This increases the number of physical qubits required per implemented logical qubit, further reducing the number of gates that can be faithfully implemented with MBQC. Cluster states using continuous variable states of photon modes have lower gate fidelities ($\approx 93\%$ for gate noise factor of 6 dB and best reported initial squeezing variance of -17 dB with respect to vacuum) \cite{Ph-MBQC} than trapped ion and superconducting qubits. However, there is no fundamental limit to the size of squeezed photon cluster states with temporal mode encoding ($\approx 10^6$ mode optical cluster state has been experimentally realized \cite{OneMil}). Quantum error correction with GKP qubit encoding \cite{GKP,Ph-MBQC} (not yet experimentally realized in the optical spectrum) could make up for the loss in individual gate fidelities and large scale deterministic quantum computation could be achieved.   

\section{\label{Conc}Conclusion}
The past decade has seen significant advances in physical realizations of qubits and the scale at which they can be coherently controlled on various platforms ranging from the most commercially deployed - superconducting and trapped ion quantum systems to those employing continuous variable states of squeezed photons which have emerged only recently. Measurement based quantum computation is an approach to build a programmable quantum computer by classically adapting single qubit measurements, based on previous outcomes, on a highly entangled cluster state comprised of a large number of qubits. In this article, we review three MBQC implementations, each utilizing a different quantum computing technology viz., superconducting qubits, trapped ion qubits and spatio-temporal modes of squeezed photons. While the latter approach has lower gate fidelities among the three, the ability to create a very large photonic cluster state using such encoding allows error correction to be employed efficiently on this platform. This could make large scale universal MBQC possible in the long run.         

\nocite{*}
\bibliography{MBQC_paper}

\end{document}